\documentstyle[12pt]{article}
\begin{document}
\thispagestyle{empty}
\vspace*{-1.5cm}
\hfill{\small KL--TH 97/2}\\[8mm]
\begin{center}
{\Large\bf Simple Calculation of Quantum Spin Tunneling Effects}\\[14mm]
{\large J.--Q. Liang}\\
{\it Department of Physics, University of Kaiserslautern, D--67653
Kaiserslautern, Germany, \\
Institute of Theoretical Physics, Shanxi University, Taiyuan, Shanxi 030 006, P.R. China,\\
Institute of Physics, Chinese Academy of Sciences, Beijing 100 080,\\ P.R. China}\\[4mm]
{\large H.J.W. M\"uller--Kirsten and J.M.S. Rana\footnote{DAAD Fellow}}\\
{\it Department of Physics, University of Kaiserslautern, D--67653\\
Kaiserslautern, Germany}\\[1cm]
{\bf Abstract}
\end{center}
The level splitting formulae much discussed in the study of spin tunneling
in macroscopic ferromagnetic particles and previously derived only by complicated
pseudoparticle methods for the ground state, are derived from those of
eigenvalues of periodic equations and extended to excited states.

\vspace{0,9cm}
\begin{center}
PACS numbers:11.15KC, 73.40GK, 75.60JP, 03.65Db
\end{center}
\newpage

In recent years the tunneling of quantum spins and the possibility of its observation in
macroscopic ferro-- or antiferromagnetic particles has attracted considerable interest
as the rapidly expanding literature in the field demonstrates\cite{1,2}.  The principal idea
is that in such particles of mesoscopic size the electronic spins can form
an aligned magnetic state which can assume several directions so that
quantum mechanics suggests the possibility to lift this degeneracy by tunneling from one
direction to another.  This tunneling effect and its possible suppression in
half--integer--spin particles (with or without applied magnetic fields) has been explored in
numerous investigations \cite{3,4,5}.  In most of these the first step is the conversion of the
discrete spin system into a continuous one \cite{6,7} by transforming spin operators
to canonical operators $\hat{p}, \hat{q}$ satisfying the quantisation relation\newline
\centerline{$[\hat{q}, \hat{p}] = \frac{i}{s}$}
where $s$ is the spin quantum number of the particle.  The calculation of the basic
semiclassical Hamiltonian is quite involved and can be found in the literature \cite{8,9},
and will therefore not be repeated here.  Once the approximate effective
quantum hamiltonian has been constructed, the next step is ususally
to proceed to the appropriate Lagrangian and then to the application
of the path--integral method.  Here, however, a problem arises.  Since the effective
Hamiltonian is that of a particle with position-- or field--dependent mass
(see below), the path--integral method must start from the phase--space
path--integral, as pointed out long ago \cite{10}. The resulting evaluation of the path--integral
by expansion about a classical configuration \cite{8,9,11,12}
which itself is already a complicated  expression 
in many cases (involving sometimes elliptic functions) is so complicated that one would 
like to have a verification of the resulting formulae, at least in some limiting cases,
although, no doubt, the pseudoparticle methods applied to these theories
are of considerable interest in themselves.  Nonetheless, particularly experimentalists
interested in spin tunneling phenomena might welcome a more straightforward 
derivation of the relevant level splitting formulae.  Our objective here is therefore
to point out that in certain cases such a simpler verification is possible
by reference to some standard (though possibly not so well--known)
quantum mechanical problems, and yields
even the splitting of the appropriate excited state levels
(a definite advantage over the instanton method).  In the following we therefore
demonstrate that in the case without an applied magnetic field the level
splitting formulae for the ground state as well as excited states
can be obtained from those of Schr\"odinger equations with periodic 
boundary conditions.  In the limit of vanishing position-- or field--dependence
of the effective mass the comparison equation is the Mathieu equation, 
and in the case with position--dependence it is the (even less well--known)
Lam$\acute{e}$ equation.  In both cases
the results are easily extendable to excited states and the range of
validity of the results can be clearly inferred from the validity of the asymptotic
solutions of these equations.  The results, which verify previous path--integral
calculations, also demonstrate that the field dependence of the effective
mass provides only a minor change.  We expect that the derivation given
here may be helpful also in the verification of the results of some other 
tunneling model theories.

The Hamiltonian of the theory describing the ferromagnetic particle is given by
$$\hat{H} = K_1{\hat{S}}^2_z - K_2{\hat{S}}^2_x$$
in terms of spin operators $\hat{S}_x, \hat{S}_z$ and $K_1 > K_2 >0$,
the two contributions describing
$XOY$--easy plane anisotropy with easy axis along the $x$ direction \cite{3}.
As mentioned above the spin operators are transformed to
canonical operators $\hat{p}, \hat{\phi}$ \cite{6} with
$ \hat{S}_z = s\hat{p}$ and $ [\hat{\phi}, \hat{p} ] = \frac{i}{s}$.
One then obtains an effective Hamiltonian which has been
used in the following form in the literature \cite{8,9,11,12}
\begin{equation}
\hat{H} = s^2\frac {{\hat{p}}^2}{2m(\hat{\phi})} + V(\hat{\phi})
\label{1}
\end{equation}
where
\begin{equation}
m(\phi) = \frac{1}{2K_1(1-\lambda\sin^2\phi)}, \;\; V(\phi) = K_2s^2\sin^2\phi
\label{2}
\end{equation}
and $-\pi\leq \phi \leq\pi$.  We consider first the case of constant mass, i.e. $\lambda = 0$,
although in actual fact \cite{8,9} $\lambda = K_2/K_1$.
The problem is to determine the splitting of asymptotically degenerate energy levels
of the Schr\"odinger equation $\hat{H}\psi = E\psi$ for the given periodic potential.  Setting
\begin{equation}
h^2 = \frac {s^2K_2}{4K_1}, \Lambda =   \frac {E-s^2K_2}{K_1} + 2h^2, b=\frac{K_1}{K_2}
\label{3}
\end{equation}
the Schr\"odinger equation becomes the Mathieu equation
\begin{equation}
{\psi}^{\prime\prime}(\phi) + \left(\Lambda - 2h^2\cos 2\phi\right)\psi = 0
\label{4}
\end{equation}
The tunneling splitting of the eigenvalues $\Lambda $ of this equation (which is twice the shift of a
single level) has been derived in the literature \cite{13} from periodic boundary
conditions. In terms of the quantum number $q_0 =
2n+1, n = 0,1,2,...$ the dominant term
can be written
\begin{equation}
\triangle_{q_0}\Lambda = \frac {2(16h)^{\frac{1}{2}q_0+1}e^{-4h}}
{(8\pi)^{\frac{1}{2}}[\frac{1}{2}(q_0-1)]!}\left(1+O(\frac{1}{h})\right)
\label{5}
\end{equation}   
For the ground state (i.e. $q_0=1$) this yields the level splitting of our original problem here, i.e.
\begin{equation}
\triangle_1E = K_1\triangle_1\Lambda= \frac{16K_1}{\sqrt\pi}\left(\frac{s}{\sqrt b}\right)^{\frac{3}{2}}
e^{-2\left(\frac{s}{\sqrt b}\right)}
\label{6}
\end{equation}
Comparing this with the result of \cite{8}  in the same limit (i.e. for no applied field and the
parameter
$a\rightarrow 0$) in \cite{8}) one obtains the same result except that in \cite{8}
the exponential factor is missing.  
The exponential factor is crucial evidence of the nonperturbative nature of the calculation
(in the calculations underlying (5) the effect of boundary conditions), and its
argument usually represents the euclidean action of the 
appropriate instanton. The exponential factor is missed in \cite{8}, presumably
as a consequence of the approximation considered which then
is the leading approximation of that under discussion and is therefore
indicative of the approximation there employed. This also shows the 
conditions on the parameters under which the result is valid.  The parameter $h^2$ of 
eq.(\ref{3}) has to be large, and the result is the dominant contribution of an asymptotic
expansion in descending powers of $h^2$.  This condition implies that $s^2 >> 4b$, as
also realised in \cite{8}. An additional advantage of the Schr\"odinger equation method
given here is that it yields immediately also the level splitting for excited states, which cannot
be obtained with the usual instanton method, but instead requires the corresponding
consideration of nonvacuum (or periodic) instantons or some other method as in \cite{12}.

We now want to take the $\phi$--dependence of the effective mass $m(\phi)$ into account.
The expression itself in eq.(\ref{2}) is reminiscent of an elliptic expression.  This suggests
searching for an analogy with an elliptic equation like the Lam$\acute{e}$ equation.  
We set $x=\sin\phi$ and $\psi=(1-x^2)^{-\frac{1}{4}}\Phi$. We then expand
the coefficient of $\Phi$ in the resulting equation
in ascending powers of $x$ and obtain an equation which
we can write (with $ -\frac{2\lambda x^2}{4} \equiv -\frac{6\lambda x^2}{4} + \lambda x^2$)
\begin{eqnarray}
& & (1-x^2)(1-\lambda x^2)\Phi^{\prime\prime} + \Bigg\{\frac{1}{4}\left[2+2\lambda 
+ x^2(3-6\lambda) +x^4(3-3\lambda)\right]  \nonumber\\
& & {} +  \left(\frac{(E-s^2K_2)}{K_1} -\frac{\lambda}{2}\right) - 
\left(\frac {s^4K_2}{\lambda K_1} -1\right)\lambda x^2  + O(x^6, \lambda^2)\Bigg\}\Phi = 0
\label{7}
\end{eqnarray}
The Lam$\acute{e}$ equation is best known in the form \cite{14}
\begin{equation}
w^{\prime\prime} + \left\{\Lambda -n(n+1)k^2 sn^2z\right\}w = 0
\label{8}
\end{equation}
The eigenvalues $\Lambda $  and eigenfunctions $w$ of the equation with periodic boundary
conditions can be found in the literature \cite{14}.  The level splitting analogous
to that of the Mathieu equation is not so well--known but
also available in the literature \cite{15}.  In fact, under
certain conditions the Lam$\acute{e}$ equation reduces to the Mathieu equation.  
In eq.(\ref{8}) $n$ is an integer in the case of Lam$\acute{e}$ polynomials, and
$k$ is the elliptic modulus of the Jacobian elliptic functions.  The quantity $\Lambda$
is the eigenvalue.  To a good approximation we can convert eq.(\ref{8}) into
one similar to eq.(\ref{7}).  To this end we set $t = sn z$ and
$w = \left[(1-t^2)(1-k^2t^2)\right]^{-\frac{1}{4}}\Xi.$
The equation for $\Xi$ is then
\begin{eqnarray}
&&(1-t^2)(1-k^2t^2)\Xi^{\prime\prime} +\Bigg\{\frac{1}{4}\left[2+2k^2
+t^2(3-6k^2) +t^4(3-3k^2)\right] \nonumber\\
&& {} + \Lambda - n(n+1)k^2t^2 +O(k^4, t^6)\Bigg\}\Xi = 0
\label{9}  
\end{eqnarray}
Comparing eq.(\ref{7}) with eq.(\ref{9}) we can identify the following
quantities
\begin{equation}
\lambda \equiv k^2, \frac {(E-s^2K_2)}{K_1} - \frac{\lambda}{2} \equiv \Lambda,
\kappa^2 \equiv \left(\frac{s^4K_2}{\lambda K_1} - 1\right)\lambda
\label{10}
\end{equation}

The level splitting of the eigenvalue $\Lambda $ of the Lam$\acute{e}$ equation is known to
be in leading order of $\kappa^2 = n(n+1)k^2$ \cite{15}
\begin{equation}
\triangle \Lambda_{q_0} =  4\kappa\left(\frac{2}{\pi}\right)^{\frac{1}{2}}
\left({\frac{1+k}{1-k}}\right)^{-\frac{\kappa}{k}}
\frac {\left(\frac{8\kappa}{1-k^2}\right)^{\frac{1}{2}q_0}}
{[\frac{1}{2}(q_0-1)]!}\left(1+O(\frac{1}{\kappa})\right)
\label{11}
\end{equation}
where for small values of $k$\newline
$$ \ln\left(\frac{1+k}{1-k}\right) = 2k + \frac{2}{3}k^3 + O(k^5).$$

In the case of the ground state ($q_0 = 1$) eq.(\ref{10}) implies the splitting of the
Schr\"odinger eigenvalue $E$ given by 
\begin{equation}
\triangle E_1 =  \frac {16K_1}{\sqrt\pi}\left(\frac{s}{\sqrt b}\right)^{\frac{3}{2}}
\frac{e^{-\frac{s}{\sqrt b} (2+\frac{2}{3}k^2) }}{\sqrt{1-k^2}}\left(1+O(\frac{1+k^2}{\kappa} )\right)
\label{12}
\end{equation}
We compare this result with that given in \cite{9} (there eq. (9a))
and calculated with the help of
a complicated instanton method.  
For the comparison we have to set in our result $k^2 = \frac{1}{b}$, so that eq.(\ref{12})
becomes
\begin{equation}
\triangle E_1 = \frac {16K_1}{\sqrt\pi}\left(\frac{s}{\sqrt b}\right)^{\frac{3}{2}}
\left(1+\frac{1}{2b}\right) e^{-\frac{s}{\sqrt b} (2+\frac{2}{3b})}
\label{13}
\end{equation}
The corresponding expression obtained in the literature \cite{9} by the instanton method
is with $B = K_1$
\begin{eqnarray}
\triangle E_1 &=&\frac {16B}{\sqrt \pi} \left(\frac{s}{\sqrt b} \right)^{\frac{3}{2}}
(1+\frac{1}{b})^{\frac{3}{4}}\left[(1+\frac{1}{b})^{\frac{1}{2}}-(\frac{1}{b})^{\frac{1}{2}}\right]
^{2s+1}\nonumber\\
&=&\frac {16B}{\sqrt \pi}\left(\frac{s}{\sqrt b}\right)^{\frac{3}{2}}(1+\frac{3}{4b})
e^{-\frac{s}{\sqrt b} (2-\frac{1}{3b})}
\label{14}
\end{eqnarray}
We observe  that except for small differences in the coefficients of the effects
of the field dependence of the effective mass, the expressions obtained by these
totally different methods agree with one another.  We interpret the small differences
in the coefficients as due to unavoidable and somewhat
different approximations in the two methods.
Our procedure here extends the previous results in several ways.  First, one can clearly
see that the result is the dominant contribution of an asymptotic expansion in
descending powers of $\kappa = \frac {s^2}{b}$.  This is therefore the parameter which is 
assumed to be large, in agreement with the condition on $h^2$ above and experimentally
relevant values.  Secondly, apart from the simple reference to the result in the 
literature, this Schr\"odinger method yields also immediately the level splitting 
of excited states which are much more difficult to obtain with a pseudoparticle method.
Since these are easily obtained from the cited formulae above we do not write
them down in detail for the problem under discussion.

Concluding we can say that the level splitting formulae, which have been much 
discussed in connection with spin tunneling in macroscopic particles but have 
previously been derived only by complicated pseudoparticle path--integral methods  
can readily be obtained in equivalently good approximation from the level
splitting of eigenvalues of periodic equations, which admittedly are not so well known
but have been studied extensively in the literature.  This simpler way of
deriving these formulae should appeal particularly to experimentalists.

\end{document}